\documentclass{aa}  
\usepackage{keyval}

\usepackage{textcomp}

\usepackage{graphicx}
\usepackage{txfonts}
\usepackage[hidelinks, colorlinks=true, linkcolor=blue, citecolor=blue, urlcolor=black]{hyperref} 

\usepackage{gensymb}
\usepackage{commath}
\usepackage{scrextend}
\usepackage{sidecap}
\usepackage{subcaption}
\usepackage{amsmath}
\usepackage{mathrsfs}
\usepackage{lineno}
\modulolinenumbers[10]

\begin{document} 

%\linenumbers
   \titlerunning{1I/'Oumuamua-like objects's SFDs with older, current and future surveys}
   
   \authorrunning{Albornoz-Montenegro et al.}
   
   \title{Constraining 1I/'Oumuamua-like objects's size distribution with older, current and future surveys}

   \subtitle{}

   %\author{R. Albornoz-Montenegrol
   %       \inst{1}
   %       \and
   %       C. Fuentes\inst{2}\fnmsep\thanks{Just to show the usage
   %       of the elements in the author field}
    %      }
    \author{Rodrigo~Albornoz-Montenegro\inst{\ref{UCH}}
    \and
          C\'esar~Fuentes\inst{\ref{UCH}}}

    \institute{Departamento de Astronom\'{\i}a, Universidad de Chile, Santiago, Chile\\
            \email{ralbornoz@das.uchile.cl} \label{UCH}}

   %\institute{Institute for Astronomy (IfA), University of Vienna,
   %           T\"urkenschanzstrasse 17, A-1180 Vienna\\
              %\email{wuchterl@amok.ast.univie.ac.at}
         %\and
         %    University of Alexandria, Department of Geography, ...\\
             %\email{c.ptolemy@hipparch.uheaven.space}
             %\thanks{The university of heaven temporarily does not
            %         accept e-mails}
            % }

% \abstract{}{}{}{}{} 
% 5 {} token are mandatory
 
  \abstract
   {}
   {Interstellar Objects (ISOs) may provide direct evidence of planetesimals ejected from early-stage solar systems. Studying these objects and predicting their detection through various surveys is vital for enhancing our understanding of planetary formation.}
   {We show how ISO size distribution (SFD), $f_{D}\!\sim\! D^{-q}$, %\old{$N(>D)\!\sim\! D^{-q}$},
   controls expected detections in current and future surveys. We focus on TESS and the Vera Rubin Observatory LSST while analyzing Pan-STARRS and ATLAS.}
   {TESS is expected to detect $\!\sim\! 1$ ISO from shallow SFDs ($q \!<\! 2.5$).
   On the other hand, LSST is likely to detect $\!\sim\! 0.2$-$100$ ISO yr$^{-1}$ over its lifetime. ATLAS is likely to detect one non-cometary object for populations with $q\!<\!3$. We further estimate $\sim\! 0$-$8$ detections of cometary like-3I/ATLAS objects by LSST in 10 years and $\sim\! 1$ cometary object detected by TESS over its mission time, independent of the nucleus size distribution. By considering both 1I/'Oumuamua and 3I/ATLAS from the same SFD, our results suggest that more asteroid-like objects are produced rather than comets during the early stages of planetary formation. Imminent detections of ISOs by LSST will help constrain the slopes of the SFDs for asteroid-like and cometary ISOs better, and, if applicable, differentiate between them.}
   {}


   \maketitle
%
%-------------------------------------------------------------------

\section{Introduction} \label{sec:intro}

The first known ISO, 1I/'Oumuamua, a rocky and elongated object with no cometary activity, with an estimated diameter and albedo of $\!\sim\!200$ m and $\!\sim\! 0.04$ respectively (\citealt{2017Natur.552..378M}), was discovered in 2017 with the Panoramic Survey Telescope and Rapid Response
System (Pan-STARRS; \citealt{2016arXiv161205560C}) survey. Pan-STARRS is a wide-field sky astronomical survey program, located at Haleakala Observatories in Hawaii. 'Oumuamua detection has involved several estimations of the galactic number density of similar bodies, which ranges between $n_{iso} \!\sim\! 0.08$-$0.2 \hspace{1mm} \mbox{au}^{-3}$ (\citealt{2017ApJ...850L..36J};
\citealt{2018MNRAS.479L..17P};
\citealt{2018}; \citealt{2019NatAs...3..594O}). Conversely, in $2019$, the second known ISO 2I/Borisov was discovered by Gennady Borisov, an engineer and amateur astronomer, as part of a near-Sun survey using a $0.65$m  f/$1.5$ astrograph (\citealt{2021SoSyR..55..124B}). 2I/Borisov showed a clear cometary activity with an extended and bright coma, and a nucleus size between ($0.2$-$0.5$) km was estimated (\citealt{2020NatAs...4...53G}; \citealt{2020ApJ...888L..23J}). Using this second detection, \citet{2021SoSyR..55..124B} computed a new constrained galactic number density of objects, with $50$ interstellar bodies $>50$ m in size in the Solar System in a sphere with a radius of $50$ au at any given time. Still, since deeper details about the survey have not been published, such depth and areal coverage, it's impossible to use the detection of 2I/Borisov to constrain further the galactic number density of these large bodies (\citealt{2023ARA&A..61..197J}).
The third and most recently discovered ISO was 3I/ATLAS\footnote{\href{https://minorplanetcenter.net/mpec/K25/K25N12.html}{\texttt{https://minorplanetcenter.net/mpec/K25/K25N12.html}}}, detected by the Asteroid Terrestrial-impact Last Alert System (ATLAS; \citealt{2018PASP..130f4505T}) on July 1, 2025. ATLAS is a sky survey system designed to find hazardous near-Earth asteroids (NEAs). Independent photometric and spectroscopic observations of 3I/ATLAS were performed immediately after the discovery and revealed cometary activity due to its red coma (\citealt{2025ApJ...989L..36S}; \citealt{2025MNRAS.544L..31O}; \citealt{2025MNRAS.542L.139B};
\citealt{Chandler_2026}). Different methods such as non-gravitational acceleration (\citealt{2025arXiv250921408C}; \citealt{Eubanks_2025}; \citealt{Forbes_2026}; \citealt{2026arXiv260315735T}; \citealt{2026ApJ...999L..37H}), nucleus extraction (\citealt{2025ApJ...990L...2J}; \citealt{2026ApJ...999L..37H}), aperture photometry (\citealt{2025ApJ...989L..36S}; \citealt{2026ApJ...999L..37H}) and surface profile comparison (\citealt{Chandler_2026}) have been used so far for 3I/ATLAS's nucleus radius constraints, in which \citet{2025ApJ...990L...2J} has set the lowest estimate $r_{n}\!>\!0.22$ km, and the higher upper limit being $r_n\!<\!10\pm 1$ km from \citet{2025ApJ...989L..36S}.

Not only meter-sized ISO has been discovered so far. Interstellar dust in our Solar System was first discovered by the Ulysses spacecraft in Jupiter in 1993 (\citealt{1993Natur.362..428G}; \citealt{1994A&A...286..915G}). After that, in 1996 \citet{1996Natur.380..323T} reported radar detection of interstellar particles in the Earth's atmosphere. Recent reports have identified potential interstellar meteors, detected by U.S. Department of Defense (DoD) satellite sensors and published by the NASA-JPL Center for Near Earth Object Studies (CNEOS). These meteors, designated IM1 and IM2 (\citealt{2022AJ....164...76P}; \citealt{2022ApJ...939...53S}; \citealt{2022ApJ...941L..28S}), have not yet had their interstellar origins fully confirmed, largely due to the lack of uncertainty data in the observations (\citealt{vaubaillon2022hyperbolicmeteorscneos20140108}). Moreover, several studies propose alternative explanations for these objects. For example, IM1 might be consistent with a typical chondritic impactor if we assume it entered the atmosphere at a lower velocity (\citealt{2023ApJ...953..167B}). Similarly, it is possible that IM2 was affected by a gravitational perturbation during the flyby of Scholz’s binary star system within the outer Oort Cloud $\!\sim\! 70$ kyr %\old{$\!\sim\! 70$ kyr$^2$} 
ago (\citealt{2014A&A...561A.113S}; \citealt{2015AJ....149..104B}; \citealt{2015ApJ...800L..17M}; \citealt{2024Icar..40815844P}). Wide reviews of the state-of-the-art of ISOs before 3I/ATLAS's discovery can be found in \citet{2023ARA&A..61..197J} and \citet{2023arXiv230317980F}.

1I/’Oumuamua source population and mechanism ejection is still unknown, and it would probably remain in the same status forever (\citealt{2023arXiv230317980F}). A lot of interpretations about their origin outside the Solar System have been proposed, from an ejection due to a close planetary encounter or a loss of mass by a compact remnant star (\citealt{2018MNRAS.479L..17P}), a molecular cloud core origin (\citealt{2021ApJ...917...20H}) or a possible ejection mechanism involving hydrogen and nitrogen ice (both cases separately) (\citealt{2020ApJ...896L...8S}; \citealt{2021ApJ...922...39L}; \citealt{2021JGRE..12606807D}), among others. As we said before, by assuming a simple power law for ’Oumuamua to attempt to constrain their size–frequency distribution (SFD), \citet{2019ApJ...884L..22R} suggested that 1I/’Oumuamua could have originated beyond the snow line, having been ejected from a DSHARP-type system (\citealt{2018ApJ...869L..41A}) via gravity assists. However, \citet{2018ApJ...866..131M} suggested that the scenario of 'Oumuamua being representative of a population of isotropically distributed objects, i.e., from a cascade collision evolution, is unlikely.

%\pt{this is what we set out to do to know it}

ISOs could be direct evidence of the ejection of planetesimals at early ages from other solar systems (\citealt{Morbidelli2003}; \citealt{GOMES2003404}), implying that the study of these objects and in particular, the prediction of a number detection by incoming new powerful surveys focused on transient events, such as Vera Rubin Observatory’s
Legacy Survey of Space and Time (LSST; \citealt{2019ApJ...873..111I}) is crucial to improve our current knowledge on the planetary formation field. LSST is a comprehensive ground-based optical survey located in Cerro Pachón, Northern Chile. During its 10-year primary mission, LSST will survey each part of the sky available in the Southern Hemisphere approximately 1,000 times. Upcoming ISOs detections by LSST seem to be imminent after the 'Oumuamua discovery by Pan-STARRS. \citet{2023MNRAS.523L...9F} predict an 'Oumuamua-like object detection with LSST in less than $5$ years with $90\%$ confidence. Previously, \citet{2019ApJ...884L..22R} predicted that LSST would detect $100$ yr$^{-1}$ objects with $r>1$m. \citet{Hoover_2022} suggested an LSST detection rate of 1-3 ISOs yr$^{-1}$ by simulating the LSST geometrical efficiency detecting objects with the same absolute magnitude as 'Oumuamua. Predictions involving different SFDs and a wide range of albedos for incoming 'Oumuamua-like objects have been performed, such as in \citet{2023PSJ.....4..230M}, where detection rates of ISOs were reported as ranging from 0-70 yr$^{-1}$. More recently, considering relistic cadence scenarios of LSST functioning, \citet{2025PSJ.....6..214D} predicted that between 6-51 ISOs are expected during LSST lifetime. While predicting how many ISOs LSST will be sensitive to is crucial, the same analysis has not been performed in current, less deep surveys, as with the Transiting Exoplanet Survey Satellite (TESS; \citealt{2015JATIS...1a4003R}). %\old{(TESS; Gangestad et al. 2013)}.

TESS is a space-based survey for planetary transits. It yields  month-long stares near the anti-solar direction. \citet{2021DPS....5350505F} showed that its limiting magnitude can be extended  \(m_{\rm lim}\!\sim\!16\) to  \(m_{\rm lim}\!\sim\!20\) using machine-learning methods. ISOs are expected to cluster near the solar apex and antiapex \citet{Hoover_2022}—as for 1I/‘Oumuamua. The long, near-continuous coverage of the ecliptic poles make TESS a useful  instrument for ISO searches (\citealt{2020PSJ.....1...81R}).

%Additionally, other current and future observing programs, primarily based in Antarctica, have this pointing advantage of searching for ISOs. Notable programs include the Antarctic Search for Transiting Exoplanets (ASTEP) (\citealt{10.1093/mnras/stad2943}) and the upcoming Cryoscope, a cryogenic infrared survey telescope to be set up in Antarctica (\citealt{2025arXiv250206950K}). 

In this work, we calculated and characterized the detection rate of ISOs of equal or smaller size than 'Oumuamua, which LSST, TESS, and ATLAS could potentially achieve, using the well-characterized detection rate of Pan-STARRS as a reference, and assuming that ISOs come from a population of dynamic origin that can be described with a simple power law for their SFDs, as do rocky bodies in the Solar System. As we expected and as other authors have done, we have shown that the number of detections per year and per campaign of ISOs depends mainly on the slope of the SFDs and, considering the uncertainties, we can constrain the population of origin and the galactic number density of ISOs through both LSST, TESS, and ATLAS detections.

%\pt{this is what follows}

This paper is structured as follows: Section \ref{sec:2} details the methodology and formalism used in this work, in Section \ref{sec:3} we present our predicted detection number for each survey, and in Section \ref{sec:4}, we discuss the implications of our results and conclude.

%the results about the number detection rates obtained by each survey and its implications, and in Section \ref{sec:5}, we conclude.

\section{Method} \label{sec:2}

We explore how the expected number of ISO detections by a survey depends on the ISO population's SFD slope. These estimates consider the spatial density of ISOs that yield a single detection ('Oumuamua) for the Pan-STARRS survey (\citealt{2018}).

%We followed the observability of each synthetic population over a short observation window (15-30 days), and computed the probability that a detection had been detected in the previous window. This approach allowed us to estimate the total number of ISOs detected for a given survey's sky coverage, and length of operation.

We followed the same formalism than \citet{2023PSJ.....4..128G} to estimate the total number of objects brighter than a certain limiting magnitude $m_{lim}$ per steradian, $\Sigma(m_{lim})$:

\begin{equation}\label{eq:1}
\Sigma(m_{lim})= \int^{1}_{0}\int^{\infty}_{0} \int^{\infty}_{D_{min}(m_{lim}, r, p)} f_{r}f_{p}f_{D}r^{2}dDdrdp
\end{equation}

Where $f_{r}r^2$, $f_p$, and $f_D$ are the heliocentric distance, albedo, and size distributions of the objects. $D_{min}$ is the minimum size for which an object with a certain position and albedo can be detected by a survey with a limiting magnitude $m_{lim}$ (Eq.$~8$ in \citealt{2023PSJ.....4..128G}). We assume $f_D$ is described by a power law with slope $q$ ($f_D \propto D^{-q}$). We assume a population of rocky and cometary objects with a constant spatial density number of ISOs, which is fixed to $f_{r} = n_{iso} = 0.1$ au$^{-3}$ (neglecting gravitational focusing effects) for ISOs greater than or equal in size to 'Oumuamua (\citealt{2019NatAs...3..594O}).

We focus on the effect of distance and size distribution on the expected ISO detections. For this reason, we ignore any
albedo distribution in our analysis. We considered two distinct values of albedos, one representative of Kuiper-belt objects $p=0.04$
(\citealt{2017Natur.552..378M}; \citealt{2012A&A...541A..93M}), and an extreme albedo of $p=0.4$.

We estimate the fraction of objects at each heliocentric distance that fall within the survey's field of view, and that are bright enough for the survey to detect. We do so by considering a synthetic population at a range of heliocentric radii, computed a velocity distribution for ISOs at those distances, and then integrating their orbits for 15-30 days and select those that appear in the field of view for at least three days (Section \ref{subsec:surveys}). The actual number of objects bright enough to be detected was determined by the minimum value of $D_{min}$ (Eq.~\ref{eq:1}) along the observed arc, taking into account the distance to the observer $\Delta$ and phase angle $\alpha$ values (Section \ref{subsub:weights}); and by considering a standard asteroidal phase function with $G = 0.15$ (\citealt{2010Icar..209..542M}; \citealt{2023PSJ.....4..124L}). Given the spatial density of ISOs larger than 'Oumaumua, $n_{iso}$, we compute the total number of new detections per observation period in a survey by numerically integrating the fraction of detectable objects over heliocentric distance (each of the synthetic populations starting at different heliocentric radii, with the maximum as $R_{\text{max}}$):

\begin{equation} \label{eq:rate1}
   N_{t}(\le m_{lim}, q) = 4\pi \int^{R_{\text{max}}}_{0 \hspace{.4mm}\text{au}} n_{r}(\le m_{lim}, q) \hspace{.6mm} r^2  dr
\end{equation}

where $n_{r}(\le\!m_{lim}, q)$ represents the spatial number density of ISOs with a characteristic SFD slope $q$ that are detectable and fall within the field of view of a survey with a given limiting magnitude $m_{lim}$. We consider the number of new detections in an observation period $N_t$ to compute the total number of discoveries over the survey's lifetime.

We consider a size limit of $10$ m for objects in our simulations. This allows us to compare our results with those of other authors under similar conditions (\citealt{2025PSJ.....6..214D}; \citealt{2023PSJ.....4..230M}). To compare our results with \citet{Hoover_2022}, we also consider 'Oumuamua's as a size limit ($\!D_{Oum}\!=\!200$ m; \citealt{2017Natur.552..378M}).

We also set to estimate the expected number of cometary-like detections. Using Hubble Space Telescope (HST), \citet{2025ApJ...990L...2J} infer \(r_n\!\sim\!0.22\)–\(2.8\) km, depending on sublimation assumptions. The object was discovered at \(r\!\approx\!4.4\) au and \(\Delta\!\approx\!3.4\) au. An inactive \(r_n\!=\!0.22\) km object has \(m\!\sim\!21.4\), \(\approx\!4\) mag fainter than its brightness \(m\!\simeq\!17\)–$18$. Because activity modeling is uncertain, we adopt a simple prescription: objects within \(r\!\le\!4.5\) au are considered to be $4$ magnitudes brighter than their bare nucleus. Additionally, we consider a cutoff size equal to the lower limit inferred for 3I/ATLAS, $D_{3\mathrm{I/A}}\!=\!0.44$ km.

\begin{figure}[h]
    %\centering
    \includegraphics[width=0.5\textwidth]{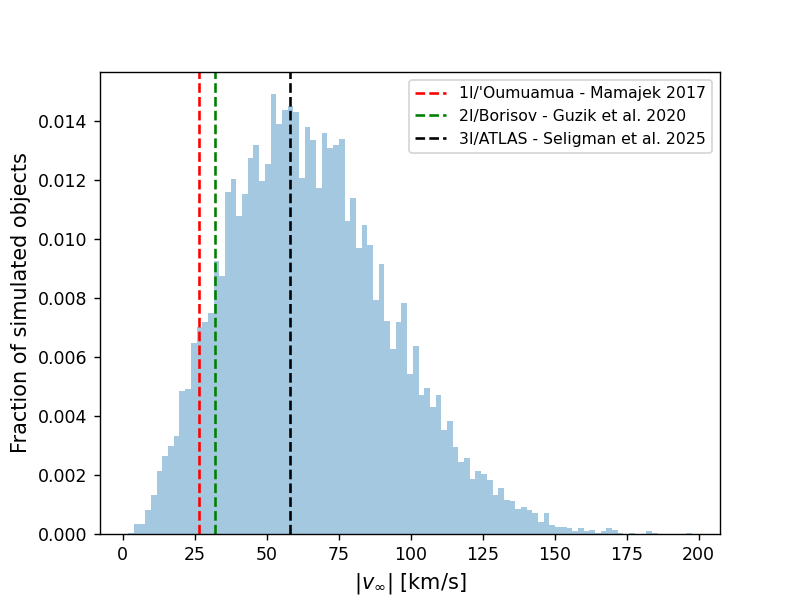}
    \caption{Distribution of the hyperbolic excess velocities $|v_{\infty}|$ for our synthetic ISOs. The red, green, and black dashed lines indicate the $|v_{\infty}|$ of 1I/'Oumuamua, 2I/Borisov, and 3I/ATLAS respectively (\citealt{2017RNAAS...1...21M}; \citealt{2020NatAs...4...53G}; \citealt{2025ApJ...989L..36S}).}
    \label{fig:max_vel}
\end{figure}

\subsection{Numerical simulations} \label{sec:style}

For our synthetic population of ISO orbits there were 14 heliocentric radii considered, evenly spaced between 0.5 and 7 au. For each case, we assumed an isotropic distribution of $10^4$ objects over that sphere.

%For each case, we assumed an \ra{Figure 1 and the associated discussion are confusingly presented. The text claims a vinf is picked from the Maxwellian distribution shown in Fig. 1, associated with a (1D) velocity dispersion of $40$ km/s. Then the Sun's velocity is subtracted. However, you cannot subtract a vector from a scalar. Moreover, the thing which is approximately a Maxwellian with a $40$ km/s dispersion is not vinf, but rather the velocity distribution of stars in the local standard of rest. The distribution of vinf, which is the magnitude of the velocity of the ISO when it encounters the Solar System is NOT Maxwellian because the Sun's velocity (vector) needs to be subtracted from the velocity (vector) of the incoming ISO. The incoming ISO's velocity vector could be crudely approximated as a 3D Gaussian with $40$ km/s dispersions (resulting in a Maxwellian speed distribution in the LSR), but it is not obvious that the authors have done this or something else. Moreover, if the authors are in fact assuming "an incoming isotropic distribution" in a heliocentric frame as they state, they must not be taking the Sun's motion into account (since this makes the incoming velocity distribution anisotropic)} \cf{Change "incoming" for ISO? distance?} isotropic distribution of $10^4$ objects over that sphere.

Each hyperbolic orbit was parameterized by an impact parameter $b$ and an initial velocity of each object from infinity $v_{\infty}$, where $|v_{\infty}|$ represents the excess hyperbolic velocity. The incoming velocity vector is constructed from a 3D Gaussian distribution with a velocity dispersion of 40 km s$^{-1}$, which is the approximate characteristic velocity distribution of stars in the Solar neighborhood (\citealt{Binney2008-ga}; \citealt{2016}). The Sun's velocity is then subtracted $\vec v_{uvw}\! = \!(11.1, 12.2, 7.3 )$ km/s  (\citealt{2010}). The resulting $|v_{\infty}|$ distribution is shown in Fig~\ref{fig:max_vel}. Since we do not consider gravitational focusing, we draw $b$ from a $p(b)\!\sim\!b$ distribution. For a uniform background density without gravitational focusing, ISOs encounters crossing the Solar System can be approximated to stars crossing the galaxy (\citealt{Binney2008-ga}), and therefore proportional to $2\pi bdb$. To randomize the initial position of the object over the plane that is normal to the velocity vector, we selected a random angle $\psi$ from a uniform distribution between ($0$, $2\pi$) (see Fig. 2 from \citealt{Hoover_2022}).

We followed the parameterization of the initial positions from \citet{Hoover_2022}. Finally, the position vector $\vec R$, which denotes the initial position, can be written as $\vec R = \alpha \vec R_{sim} + \vec b$, where the $\alpha$ is the positive factor that satisfies $R_{sim} = |\alpha \vec R_{sim} + \vec b|$. The velocity at $R_{sim}$ is larger than $v_{\infty}$ by a factor of $\sqrt{(1+v_{\text{esc}}^2/v_{\infty}^2)}$ due to the Sun's gravity, so we multiply our constructed velocity vector by this factor, %\old{We only correct for the change in velocity magnitude due to gravitational focusing, we multiply the constructed velocity vector by $\sqrt{1 + v_{esc}^2/v_{\infty}^2}$},
where $|v_{esc}|$ is the escape velocity from an object at $R_{sim}$. We note that we do not account for the gravitational focusing propagation, which preferentially increases the flux of ISOs with low heliocentric velocities.

The objects are modeled forward using the IAS15 N-body integrator in \verb|REBOUND| (\citealt{2012}) for 30 days, and their positions saved every 3 days.

\subsection{Surveys}
\label{subsec:surveys}
%\pt{surveys vary by this and I modelled them according to this }
Surveys are defined by their limiting magnitude $m_{lim}$, their effective area \(\Omega\), and the timing between observations. We assume the cadence in each survey is effective in detecting ISOs. An object is observed if it is in the field of view of the survey and brighter than $m_{lim}$ in at least two different positions saved in the 3-day intervals. Although we don't directly take into account trailing loss as a factor affecting the detection of objects relatively closer to the observer, we considered the limiting magnitudes reported for each survey with respect to Solar System objects, mostly asteroids. % criteria for objects. Trailing loss is influenced by the apparent motion of the objects, which primarily depends on their distance. Instead, we have chosen to focus on the limiting magnitudes reported for each survey concerning Solar System objects, mostly asteroids.

%\ra{In line 200, could the authors clarify what exactly is meant by 2 separate positions? Are these positions the saved 3-day intervals? Why 2 positions? Typically more observations are necessary to constrain an object's orbit or even verify that it is the same object across multiple images}

\subsubsection{TESS}
%\pt{why is this special, and properties}
AI-aided methods (\citealt{2026AJ....172...34P}) or the simple co-addition of multiple frames (\citealt{2021DPS....5350505F}) can detect significantly fainter moving objects than the single-frame magnitude limit in several surveys. For TESS, we consider $m_{lim}\! = \!20$ (\citealt{2021DPS....5350505F}) for the asteroid-like ISO analysis, while for the cometary-like objects we consider the magnitude at which 3I-ATLAS was pre-discovered: $m_{lim}\! = \!21.63$ (\citealt{2025ApJ...991L...2F}). We consider a $24\times96$ deg$^2$ field of view, covered by four cameras (\citealt{2015JATIS...1a4003R}). The field of view extends from 6 deg off the ecliptic to the ecliptic pole. Every 27 days, the survey points to an adjacent pointing, covering each hemisphere in 13 sectors. We keep track of objects that move from one sector to the next within a hemisphere. We consider 8 years of TESS time mission to perform the analysis.

\subsubsection{LSST} \label{subsec:222}

LSST features an 8.4-meter primary mirror and a 3.2-gigapixel camera, with %allowing it to observe a 
9.6 deg$^2$ field of view (\citealt{2019ApJ...873..111I}). LSST will capture images in six filters: $ugrizy$, covering the wavelength range of $320$–$1050$ nm. A single-visit depth in the $r$-band filter is expected to reach $r \!\sim\!24.5$ (\citealt{2019ApJ...873..111I}). Therefore, we consider this limiting magnitude $m_{lim}\! = \!24.5$ for LSST estimations. We consider an approximation of the typical 15-day sky coverage\footnote{\href{https://github.com/rhiannonlynne/notebooks/blob/main/Template\%20Generation.ipynb}{\texttt{https://github.com/rhiannonlynne}}} of the LSST cadence
simulation \verb|baseline_v3.0_10yrs| (\citealt{2023...zenodo}; \citealt{2019AJ....157..151N}), as shown in Fig.~\ref{fig:all_sky}. ISO's heliocentric coordinates are transformed into celestial coordinates. At each time of the 3-day time-spaced orbit, we compute the local coordinates of each object from LSST during that specific day with a resolution time of 2 hours. We consider the object detectable by LSST during this period if the ISO reaches all of the following conditions:

\begin{itemize}
    \item The ISO's celestial coordinates, right ascension and declination, are within the field of view of LSST shown in Fig.~\ref{fig:all_sky} at least once during the 15-day observing period.
    \item The ISO altitude is higher than 30 deg, as seen from the LSST location, and the Sun's altitude is lower than -18 deg. This last ensures a before/after astronomical twilight observation.
\end{itemize}

We repeat this process for the rest of the orbit, another $15$-day period, and if the same object reaches the above conditions, we consider it a repeating object. This time interval was chosen to be sufficiently long to estimate the number of objects that the same survey might rediscover. %\ra{In section 2.2.2 could the authors please clarify where the choice of 15 days originates?}

\subsubsection{ATLAS}
\label{subsec:223}
ATLAS comprises four independent robotic telescopes located around the world: two in the Northern Hemisphere at the Haleakala (MPC code T05) and Mauna Loa (MPC code T08) observatories in the Hawaiian Islands, and two in the Southern Hemisphere at the El Sauce observatory (MPC code W68) in Chile and the Sutherland observatory (MPC code M22) in South Africa. Each unit consists of a 0.5m Schmidt telescope, a 0.65m spherical primary mirror, and a camera STA 1600 10,560 × 10,560 pixel single CCD
detector with a 29 deg$^2$ field of view (\citealt{2018PASP..130f4505T}; \citealt{2021PASP..133b9201S}).

%3I/ATLAS was discovered using the o-band filter, but for generality, 

3I/ATLAS was discovered by the ATLAS-Chile telescope (which we will refer to by its MPC code, "ATLAS-W68"). This unit began operations in early 2022, and as of now, we consider a mission duration of 3.5 years. 
ATLAS uses two main filters: the $c$-band filter ($420$–$650$nm) for darker nights and the $o$-band filter ($560$–$820$nm) for brighter nights. Both have limiting magnitudes of $\!\sim\!19.5$ and $\!\sim\!19$ respectively (\citealt{2018PASP..130f4505T}; \citealt{2021PASP..133b9201S}).
We use a limit $m_{lim}\! = \!19.5$ and consider a 15-day sky coverage of ATLAS-W68 during which 3I/ATLAS was observed, specifically from June 24 to July 9, 2025. We obtained the raw data pointing from the pointing data service of the Minor Planet Center\footnote{\href{https://minorplanetcenter.net/mpcops/pointings/sky_coverage/}{\texttt{https://minorplanetcenter.net/mpcops/pointings/}}}. The total sky area observed during this time was $\!\sim\!12$,$100$ deg$^{2}$, and we estimate and utilize an equivalent area as shown in Fig.~\ref{fig:all_sky}. Detection criteria for the ISO's objects are analogous to those described for LSST. We assume that the same sky area sampled is the same for the subsequent 15 days. Therefore, we can apply the same method used for LSST to identify the repeating objects during each 15-day observation period.

\subsubsection{Pan-STARRS} \label{subsec:224}
Pan-STARRS primary facility, PS1, features a 1.8-meter telescope equipped with a 1.4-gigapixel camera, capturing images in the $grizy_{P1}$ broadband filters (\citealt{2016arXiv161205560C}). \citet{2018} estimated the volume surveyed by Pan-STARRS during a typical 80-day period of observations, including the night when 'Oumuamua was discovered (MJD 58045) using the \( w_{P1} \)-band filter ($400$-$800$nm), which integrates the \( gri \)-bands %and is optimized for the Solar System Survey segment of the mission 
(\citealt{2016AJ....152..147L}), with $m_{lim}\! = \!22$. 

During these 80 days, a total volume of $0.3$ au$^3$ was surveyed (\citealt{2018}). This suggests that approximately $\!\sim\!9.2$ au$^3$ is observed over $\!\sim\!6.75$ years of time mission. A single detection during this time implies a mean density of $\!\sim\! 0.1$ au$^{-3}$ for ISOs larger than or equal in size to 'Oumuamua.

%\cfg{acá}
We consider a 15-day sky-coverage of Pan-STARRS during the month in which 'Oumuamua was discovered, within the range from MJD 58041-58055 (raw data pointings were obtained through private communication with Pan-STARRS staff). Some nights within this range were missing due to the bad weather, but we will ignore them. The total area observed during this routine was approximately $4,500$ deg$^2$. An equivalent sky area surveyed during this time is shown in Fig.~\ref{fig:all_sky}. Detection criteria, sky area sampling ratios, and treatments for repeating objects are analogous to those used in the LSST and ATLAS procedures.

\subsubsection{Optimized follow-up observations} \label{subsub:weights}
To optimize computational efficiency at the beginning of each ISO trajectory, we assume that each survey shares the same heliocentric longitude as the object, denoted as $\phi_{hel, 0}$ for the specific ISO. In the case of TESS, this can be done directly since TESS always points to the anti-solar direction, and sunlight does not represent a problem. However, this is not true for LSST and Pan-STARRS. 

We determined the specific date for each heliocentric position of the Earth throughout 2025 (the choice of year is arbitrary and does not significantly impact the analysis) with a resolution of 1 day, using an ephemerides dataset accessible through the Python package \verb|SOLARSYSTEM| (\citealt{Nasios:2020}). For the subsequent days of the ISO orbit, we consider the dynamics of each survey moving as a uniform circular motion (UCM) at $v_{\perp}\! = \!30$ km/s. By considering both the survey and ISO positions across the 15-30 day orbit, we can calculate the distance to the observer and the phase angle at any point in ISO's orbit, which are used in Section \ref{sec:2}, to determine the minimum visible $D_{min}$ of the ISO during its orbit.

Given the ideal scenario assumed for each interstellar object at the start of its trajectory, we have to geometrically correct them, as this scenario is often not realistic.

The subtended arc in longitude traveled by some object "$j\hspace{.4mm}$" between the discretized moment $n-1$ and $n$, can be denoted as $l_{n-1, n}^{j}$. Therefore, the total arc traveled by the object during its trajectory will be the sum of these sub-arcs. One can consider the total change in longitude (in radians) traveled by the object during its trajectory, denoted as $|\Delta \phi_{hel}^{j}|$. This change is weighted by the heliocentric latitude at which the object started its trajectory, represented as $\theta_{hel, 0}^{j}$, which is defined within the range: $-90$° $\le \theta_{hel, 0}^{j} \le 90$°. As a result, the effective arc traveled by the object $j$ during its trajectory can be expressed as:

\begin{equation}
    l_{eff}^{j} = |\Delta \phi_{hel}^{j}| \hspace{0.5mm}\mbox{cos}(\theta_{hel, 0}^{j})
\end{equation}

Therefore, the geometric correction for the object $j$ to be within the field of view of the survey is:

\begin{equation} \label{eq:omega}
    \omega^{j} = \text{min}\hspace{.5mm}\Bigg(\frac{L + l_{eff}^{j}}{2\pi\hspace{0.5mm}\mbox{cos}(\theta_{hel, 0}^{j})}, \hspace{1mm}1\Bigg)
\end{equation}

%\old{
%\begin{equation} \label{eq:omega_old}
%    \omega^{j} = \frac{L + l_{eff}^{j}}{2\pi\hspace{0.5mm}\mbox{cos}(\theta_{hel, 0}^{j})}
%\end{equation}
%}

where $L$ (in radians) is the sky-projected size in the camera's longitude, and $\theta_{hel, 0}^{j}$ is the initial heliocentric latitude of the object. The definition of Eq.~\ref{eq:omega} is because there are cases where $\omega^{j}\! > \!1$. For instance, with TESS, this occurs when $|\theta_{hel, 0}^{j}|\! > \!78$°, since the camera covering the entire ecliptic pole captures every object starting at any heliocentric latitude $\ge\! \!78$°. %In fact, as $|\theta_{hel, 0}^{j}| \!\rightarrow\! 90$°, $\omega^{j} \!\rightarrow\! \infty$.}

The weight $\omega^j$, therefore, is multiplied by the actual number of objects bright enough to be detected in each simulation as a function of $r$, and the fraction of visible objects is calculated by averaging this number with the total number of particles in each simulation, computing an ''detection efficiency'' as a function of $r$.

%\old{Should be noticed that there are cases where $\omega^{j}\! > \!1$. For instance, with TESS, this occurs when $|\theta_{hel, 0}^{j}|\! > \!78$°, since the camera covering the entire ecliptic pole captures every object starting at any heliocentric latitude $\ge\! \!78$°. In fact, as $|\theta_{hel, 0}^{j}| \!\rightarrow\! 90$°, $\omega^{j} \!\rightarrow\! \infty$ . Therefore, for values where $\omega^{j} \!>\! 1$, we set $\omega^{j}\! = \!1$.}

%\subsection{Number of Detections}

%To compute the total number of ISOs observed by a survey over a given period of observation, we consider the detection efficiency from each set of simulated orbits at different heliocentric distances $r$, therefore Eq.~\ref{eq:1} is integrated over the solid angle and we obtain:

%\begin{equation} \label{eq:rate1}
%   N_{t}(\le m_{lim}, q) = 4\pi \int^{R_{\text{max}}}_{0 \hspace{.4mm}\text{au}} n_{r} \hspace{.3mm} r^2  dr
%\end{equation}

%where $n_r$ is the expected density number of objects (given a constant density $n_{iso}$ for ISOs larger to ’Oumuamua) brighter than a survey's limiting magnitude $m_{lim}$, and visible by the survey (as is detailed in Section \ref{subsec:surveys}):

\begin{equation} \label{eq:eta_r}
   n_{r} = \chi_{r}\times \sum^{S}_{j} \Bigg (  \frac{\omega_r^{j}}{S}\Bigg ) \times  n_{iso}\times \Bigg(\frac{D_{min}^j}{D_{Oum}}\Bigg)^{1-q} 
\end{equation}

%where $\omega_r^{j}$ is defined in Eq.~\ref{eq:omega} for each ''j'' object. $S$ is the total number of synthetic particles in each simulation ($10^4$) and $\chi_{r}$ is the fraction of the visible objects that are not repeated in a subsequent observation period, i.e., a new object. Finally, the number of ISOs detected per campaign for each survey is given by Eq.~\ref{eq:Nq}:

%\begin{equation} \label{eq:Nq}
%N\hspace{.4mm}(\le m_{lim}, q) = N_t\hspace{.4mm}(\le m_{lim}, q) \times \kappa \times 12 \times T
%\end{equation}

%where $\kappa$ is a factor to account detection rates monthly and $\kappa \!=\! 2$ for LSST, Pan-STARRS and ATLAS-W68, since their simulated observation periods were 15 days. For TESS, $\kappa \!=\! 1$. $T$ is the total length time of the survey in years.

%In other words, we are sampling the detection efficiency of each survey together with the expected number of objects present given a size, distance, and albedo distribution, per unit distance. Finally, the total number of objects detectable by the survey is obtained by integrating the above quantity over volume, and then weighted consistently to translate to monthly, yearly, and campaign detections.

\section{Results} \label{sec:3}

\begin{figure}[h!]
    \centering
    \includegraphics[width=0.49\textwidth]{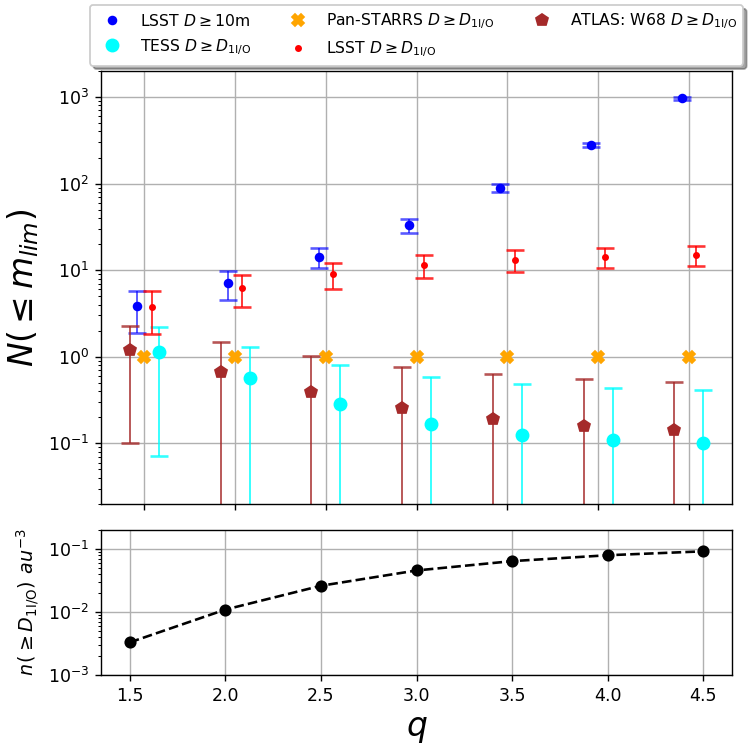}
    \caption{\textbf{(top)} Number of asteroid-like ISOs per campaign detectable by each survey for different SFD slopes and their respective uncertainties. Yellow crosses are fixed at one because they represent the empirical rate of detection per operation time in which 'Oumuamua was discovered by Pan-STARRS (\citealt{2018}). \textbf{(bottom)} The spatial number density of asteroid-like ISOs greater than or equal in size to 'Oumuamua constrained by the number of detections from Pan-STARRS, as a function of their SFD. Small offsets in $q$ were added for clarity, where the range of $q$ spans from $1.5$ to $4.5$, with increments of $0.5$ evenly spaced.}\label{fig:final_num}
\end{figure}

The expected detection number of rocky objects greater or equal in size than $10$ m and $D_{oum}$, of each survey during its lifetime mission (assuming an albedo of $p\!=\!0.04$), scaled by the empirical number detection done by Pan-STARRS, is shown in Fig. \ref{fig:final_num}. The uncertainties on our estimates do not consider the uncertainty on the spatial density of objects in Fig. \ref{fig:final_num}. The uncertainties of our detection rates are shown, which are estimated as Poisson noise ($\sqrt{N}$). TESS and ATLAS detections are quite similar when using both cutoff sizes, so the $\geq\!10$ m case is shown.

TESS is more sensitive to ISOs population with shallower SFD, potentially allowing the detection of $\!\sim\! 1$ object for $q\!<\!2.5$ during the mission. LSST would have a different detection rate depending on the slope $q$ of the SFD of the ISOs, being more sensitive to ISO populations with steeper SFD. A number of $\!\sim\!0.2$-$100$ ISO yr$^{-1}$ and $\!\sim\!0.2$-$2$ ISO yr$^{-1}$ for objects $D\!\geq\!10$ m and $D\!\geq\!D_{Oum}$ respectively, are expected during LSST lifetime mission. Finally, we predict that, within the uncertainties and based on the mission lifetime of ATLAS-W68 thus far, it's capable of detecting a single ISO originating from a source population characterized by $q\!<\!3$ (Fig.~\ref{fig:final_num}).

\begin{figure}[h!]
    %\centering
    \includegraphics[width=0.49\textwidth]{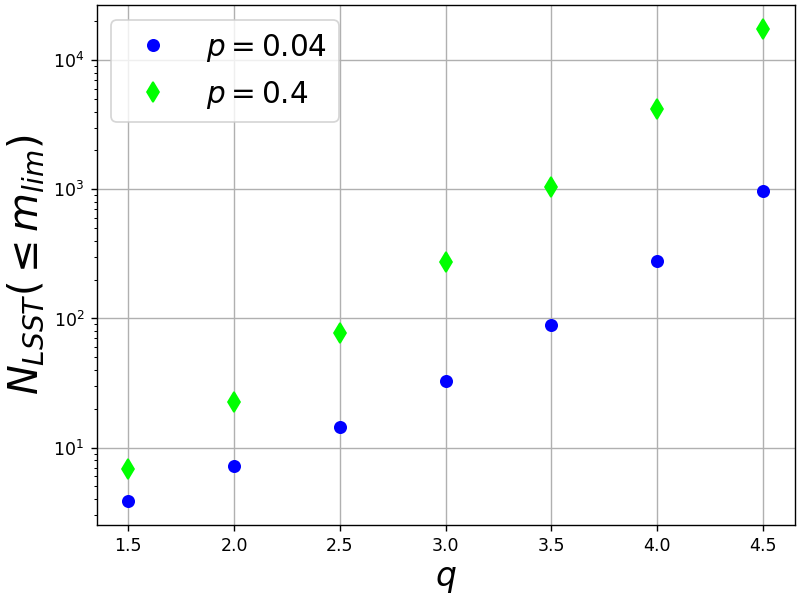}
    \caption{Detection rate of asteroid-like ISOs by LSST during their mission lifetime within their uncertainties, assuming two extreme albedos. Uncertainties are not included to make the Figure more comprehensible.}
    \label{fig:lsst_mission_p}
\end{figure}

The detection rate of asteroid-like ISOs ($D\!\geq\!10$ m) by LSST during 10 years is shown in more detail in Fig. \ref{fig:lsst_mission_p}, using different SFDs slopes and with two extreme albedos. Both assumptions about the albedo resulted in a substantial change in the final number, ranging between $\!\sim\! 0.2$ and $1$,$800$ ISO yr$^{-1}$.

\begin{figure}[h!]
    \centering
    \includegraphics[width=0.5\textwidth]{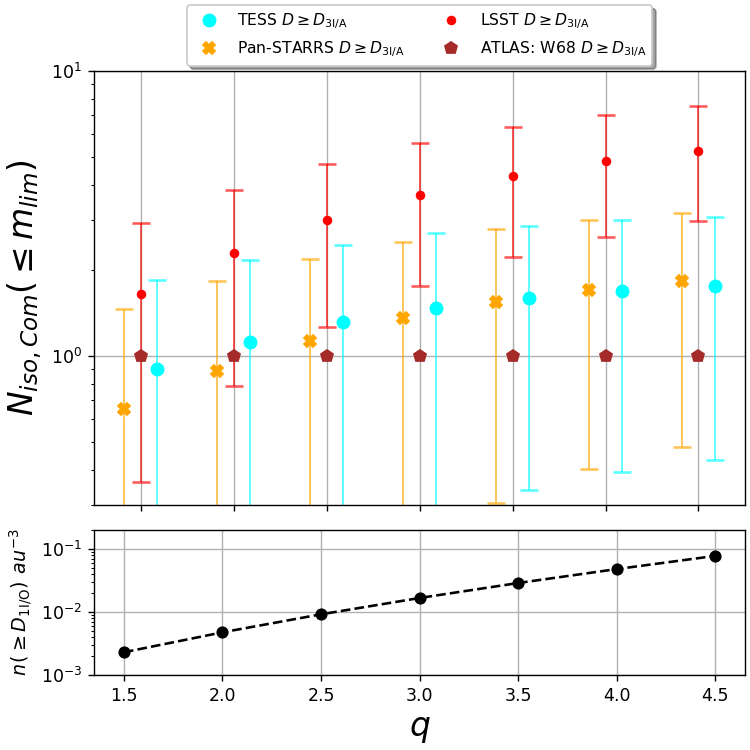}
    \caption{\textbf{(top)} Number of cometary-like ISOs per campaign detectable by each survey for different SFD slopes and their respective uncertainties. Brown pentagons are fixed at one because they represent the empirical detection rate per operation time, in which ATLAS-W68 discovered 3I/ATLAS. Small offsets in $q$ were added for clarity. The $q$ range is the same as in Fig. \ref{fig:final_num}.
    \textbf{(bottom)} The spatial number density of cometary-like ISOs greater than or equal in size to 'Oumuamua constrained by the number of detections from ATLAS-W68, as a function of their SFD.}
    \label{fig:finalCom}
\end{figure}

\begin{figure*}[h]
    \centering
    \includegraphics[width=1.05\textwidth]{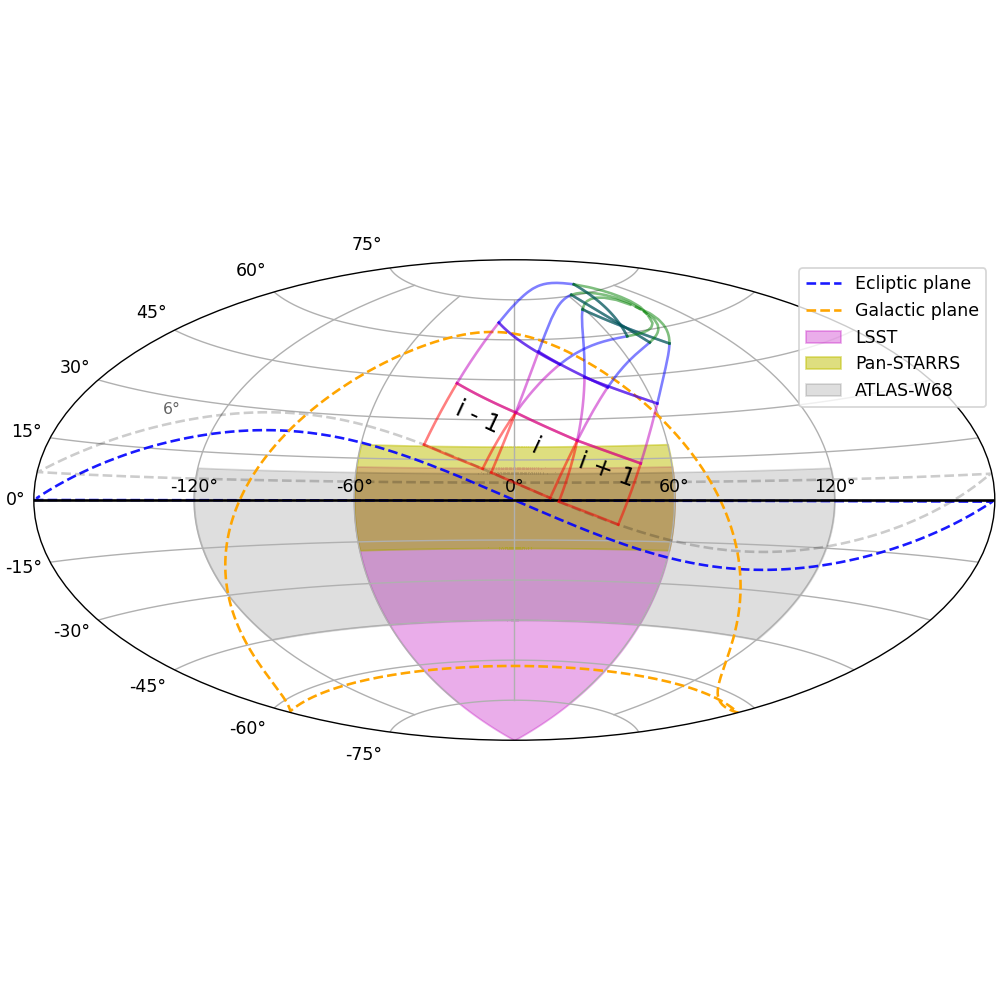}
    \caption{Sky-coverage of the four surveys utilized in this study, with celestial coordinates displayed. It depicts three consecutive observation sectors of TESS, labeled as $i\!-\!1$, $i$, and $i\!+\!1$. Each of the four TESS cameras is represented by a different color. The beginning of each sector is marked in gray at an ecliptic latitude of $6$ deg. The green area indicates the observing region of the fourth camera, which begins at an ecliptic latitude of $78$ deg and encompasses the entire ecliptic pole. The magenta, yellow, and grey areas depict the typical 15-day observation coverage of LSST, Pan-STARRS, and ATLAS-W68, respectively.
    }
    \label{fig:all_sky}
\end{figure*}

Assuming one ATLAS-W68 comet detection, Fig.~\ref{fig:finalCom} shows the expected ISO comet detection. Pan-STARRS and TESS are expected to have detected $0$-$3$ such comets, while LSST should detect $\sim\! 0$-$8$ cometary ISOs larger or equal in size to $D_{3\mathrm{I/A}}$ in 10 years.

Since the expected number of detections is linearly proportional to the galactic number density of objects, we can estimate new constraints on the galactic number density, $n_{iso}\!=\!0.1$ au$^{-3}$, for both asteroid- and cometary-like ISOs, by scaling with the detection rate from Pan-STARRS and ATLAS-W68, respectively. The bottom plot in Figs.~\ref{fig:final_num} and \ref{fig:finalCom} illustrates an updated galactic number density of objects equal to or larger than 'Oumuamua, constrained by the number of detections from Pan-STARRS and ATLAS-W68, respectively, as a function of their size distribution.

\section{Discussion} \label{sec:4}

Using the formalism introduced in \citet{2023PSJ.....4..128G} that accounts for size, distance and albedo, we estimated the ISO yield of TESS, LSST and ATLAS-W68 over their lifetime.

With its sky coverage, and by extending its nominal limit to $V\!\sim\!20$ limit, TESS is likely to detect an ISO for shallow size-frequency slopes ($q\!<\!2.5$) (Fig.~\ref{fig:final_num}). 

The deeper $r\!\sim\!24.5$ limit of LSST translates to $\!\sim\!0.2$-$100$ ISOs yr$^{-1}$ for a typical albedo $p\! = \!0.04$ and up to $\!\sim\! 0.2$-$1$,$800$ ISO yr$^{-1}$ for the $p\! = \!0.4$ case (Fig.~\ref{fig:lsst_mission_p}). Cosmic-ray radiation in the ISM is expected to darken ISOs from the typically higher than $p\! = \!0.28$ of icy TNOs (\citealt{2012A&A...541A..93M}) to a lower $p\! = \!0.04$ (\citealt{2017Natur.552..378M}). We expect LSST to detect more ISOs if the source population's size distribution is steeper, yielding a growing number of ISOs smaller than Oumuamua (Fig.~\ref{fig:final_num}). The spatial number density $n_{iso}$ for asteroids seems to be consistent with a steeper SFD characterized by $q \!>\! 4$ (Fig.~\ref{fig:final_num}), being aligned with the steep size distribution index inferred in \citet{2017ApJ...850L..36J}.

Our estimate of the detection numbers by LSST, TESS and ATLAS-W68 in Fig.~\ref{fig:final_num} is strongly a result of the assumption that Pan-STARRS was expected to observe exactly 1 ISO. Qualitatively, this can be interpreted as follows: for TESS and ATLAS-W68, given its low limiting magnitude, the survey is less sensitive to populations with steeper size distributions than Pan-STARRS. Conversely, with LSST, the situation is different. Its deeper limiting magnitude makes it more sensitive to populations with steeper size distributions than Pan-STARRs, as the difference in the number of smaller and larger objects becomes more pronounced.

There are several differences between our results and assumptions about LSST detections (Fig. \ref{fig:final_num} and \ref{fig:lsst_mission_p}) compared to previous estimates. For instance, \citet{Hoover_2022} predicted a detection rate of approximately $1$-$3$ ISO yr$^{-1}$, based on the assumption that all objects had the same absolute magnitude as 'Oumuamua. Their results align with our estimates for objects that are equal to or larger than 'Oumuamua. However, they did not take into account how the varying SFD of these ISOs could influence the detection rate, particularly if a cutoff size lower than $D_{Oum}$ is used.

\citet{2023PSJ.....4..230M} explored different $q$ slopes and albedos, leading to significant variations in the number of detections. They spanned a range of $q$ slopes from 1 to 3. However, they pointed out that focusing on an albedo distribution ranging from $0.01$ to $0.4$, their detection estimates varied between $0$ and $15$ ISO yr$^{-1}$. When limiting our analysis to the same $q$ range, our results are consistent but higher than theirs, as we predict detections of $\sim \!0.2$-$30$ ISO yr$^{-1}$ (Fig.~\ref{fig:lsst_mission_p}).

On the other hand, \citet{2025PSJ.....6..214D} recently reported a potential LSST detection number of up to 51 over its lifetime mission by varying different SFDs while using a fixed albedo of \(p=0.05\). This estimate is lower than ours, as they considered trailing loss, as in \citet{2023PSJ.....4..230M}, but also LSST cadence details and LSST Solar System Processing (SSP) discovery criteria.

As illustrated in Fig.~\ref{fig:final_num}, ATLAS-W68 shows sensitivity to a shallow size distribution of objects with $q\!<\!3$ during its lifetime mission. Reports of cometary activity in 3I/ATLAS (\citealt{2025ApJ...989L..36S}; \citealt{2025MNRAS.544L..31O}; \citealt{2025MNRAS.542L.139B}; \citealt{Chandler_2026}) highlight the presence of anisotropic dust emission, so the finding of relatively smaller and brighter comet-like ISOs suggests the existence of a larger population of comets, with 'Oumuamua being an exception rather than the norm. 

Pan-STARRS' expected detections cometary-like ISO is clearly illustrated in Fig.~\ref{fig:finalCom}; notably, 2I/Borisov should have been pre-discovered by PS1 on January 2019, when the object presented weakly cometary activity and was at $r_{h}\! \sim \!7.8$ au, almost 7 months before its discovery in August of the same year; however, the object fell within a wide chip gap \citep{2020AJ....159...77Y}. For TESS, the expected detection of interstellar comets is consistent with 3I/ATLAS’s pre-discovery at $r_{h}\! \!\sim\! \!5.4$-$6.4$ au \citep{2025ApJ...991L...2F,2025ApJ...994L..51M}. This precedent motivates systematic reprocessing of TESS data for hidden ISO comets.

The spatial number density $n_{iso}$ for comets in Fig.~\ref{fig:finalCom} appears to align with a steeper SFD ($q \!>\! 4.5$) than the asteroid case (Fig.~\ref{fig:final_num}). This suggests that if both ‘Oumuamua and 3I/ATLAS belong to the same population, more asteroids than comets are being produced during the early stages of planet formation, which is contrary to what one would expect in planetesimals produced during planetary formation processes (\citealt{BRASSER201340}).

Our assumptions about the dynamical origin of the 'Oumuamua-like population might be inaccurate, as suggested by \citet{2018ApJ...866..131M}. Gaia DR3 suggests that the ISO population in the Solar neighborhood may inherit the chemodynamically substructured velocity distribution of their progenitor stars \citet{2025AJ....169...78H}, a factor also considered by \citet{Hoover_2022, 2023PSJ.....4..230M, 2025PSJ.....6..214D}. This velocity distribution variation may alter the efficiency of each survey due to trailing losses.

The same methodology applies to other small-body populations, notably Long Period Comets (LPC), whose faint absolute magnitude at large perihelia have limited the number of discoveries to date (\citealt{2012MNRAS.423.1674F}). Applying it to future LSST detections could constrain their dynamical histories (\citealt{2025Icar..42916443I}).

We expect that the total number of ISOs discoveries by LSST will constrain the slope of their size distribution. The relative number of detected comets versus asteroids should indicate if these populations are related to each other.

\bibliographystyle{aa}

\bibliography{refs}

\end{document}